\documentclass[twocolumn, prl, showpacs,superscriptaddress,floatfix]{revtex4}
\usepackage{graphicx}
\usepackage{dcolumn}
\usepackage{bm}
\usepackage{amsmath}
\usepackage{color}

\begin{document}

\title{Super-Tonks-Girardeau gas of spin-1/2 interacting fermions}
\author{Liming Guan}
\affiliation {Institute of Physics, Chinese Academy of Sciences,
Beijing 100190, China}

\author{Shu Chen}
\email{schen@aphy.iphy.ac.cn} \affiliation {Institute of Physics,
Chinese Academy of Sciences, Beijing 100190, China}

\begin{abstract}
Fermi gases confined in tight one-dimensional waveguides form two-particle bound states
of atoms in the presence of a strongly attractive interaction. Based on the exact
solution of the one-dimensional spin-1/2 interacting Fermi gas, we demonstrate that a
stable excited state with no pairing between attractive fermionic atoms can be realized
by a sudden switch of interaction from the strongly repulsive regime to strongly
attractive regime. Such a state is an exact fermionic analog of the experimentally
observed super-Tonks-Girardeau state of bosonic Cesium atoms [Science \textbf{325},
1224 (2009)] and should be possible to be observed by the experiment. The frequency of
the lowest breathing mode of the fermionic super-Tonks-Girardeau gas is calculated as a
function of the interaction strength, which could be used as a detectable signature for
the experimental observation.

\end{abstract}
\date{\today}
\maketitle

\textit{Introduction.---} Exploring new quantum phases and understanding the striking
consequences of correlation in strongly interacting atomic gases are at the frontier of
current research in condensed matter physics and cold atom physics
\cite{Bloch,Lewenstein}. In recent years, remarkable progress has been made in the
experimental realization of fundamental many-body model Hamiltonians, such as the
Hubbard model \cite{Greiner} and the Tonks-Girardeau (TG) gas \cite{Paredes,Kinoshita},
with unprecedented tunability. In general, attentions are devoted to the exotic
properties of ground states and low-excited states. A recent experimental breakthrough
is the realization of the super Tonks-Girardeau (STG) gas of bosonic Cesium atoms
\cite{Haller}, which is a stable highly excited state of interacting Bose gas
\cite{Astrakharchik1,Astrakharchik2,Batchelor}. In the experiment \cite{Haller}, a
one-dimensional (1D) Bose gas was initially prepared in the strongly repulsive TG
regime, and then the STG gas was obtained by suddenly switching the interaction from
strongly repulsive to attractive regime. A striking feature of the STG gas is its
counterintuitive stability against collapsing to its cluster ground state even in the
presence of strongly attractive interactions.

So far, the experimental study \cite{Haller} and most of the theoretical works
\cite{Astrakharchik1,Astrakharchik2,Chen,Batchelor,Girardeau_STG,Tempfli} on the STG
gas have focused on the bosonic system. In this work, we study the possible realization
of the Fermi super Tonks-Girardeau (FSTG) state in a 1D Fermi gas. As the 1D Fermi gas
with tunable interaction strengths has already been experimentally realized
\cite{Moritz}, it is promising to directly observe the STG state in 1D Fermi gases.
Stimulated by the experiment of the Bose STG gas, we first prepare a strongly repulsive
spin-1/2 Fermi gas, and then suddenly switch the interaction to the strongly attractive
regime. By this way, we can access a stable highly excited state of the attractive
Fermi gas which does not fall into its attractive ground state.
In the strongly attractive limit, atoms with different spins form tightly bound fermion
pairs \cite{C.N.Yang,Gaudin,Chen2}. It has been shown that the ground state of a 1D
attractive spin-balanced Fermi gas is effectively described by the STG state of bosonic
pairs of fermions with attractive pair-pair interaction \cite{Chen2}. The FSTG state
being studied in the present work is essentially the lowest gas-like excited state
composed of unpaired fermions which is totally different from the bosonic STG state
composed of tightly bound fermion pair proposed in Ref.\cite{Chen2}.

\textit{Interacting Fermi model.---} We consider a system of
$N=N_{\uparrow}+N_{\downarrow}$ spin-1/2 fermions in a tightly confined waveguide
described by the effective 1D Hamiltonian
\begin{equation}
H=-\frac{\hbar ^2}{2m} \sum_{i=1}^N\ \frac{\partial ^2}{\partial x_i^2}
 + g_{1d}\sum_{i<j}\delta (x_i-x_j), \label{HF}
\end{equation}
where $g_{1d}=-2\hbar ^2/(ma_{1d})=\hbar ^2c/m$ is the effective 1D interaction
strength and $a_{1D}$ the effective 1D scattering length \cite{Olshanii}.  Without loss
of generality, we assume that $N_{\downarrow} \leq N_{\uparrow}$.
The 1D interacting spin-1/2 Fermi gas is only solvable for the homogenous case
\cite{C.N.Yang}. However, in the infinitely repulsive limit, a generalization of
Bose-Fermi mapping \cite{Girardeau} to the spin-1/2 Fermi system
\cite{Girardeau07,Guan} allows us to construct analytically exact solution of 1D Fermi
gases even in trap potentials. In this work, we focus on the homogenous system which
can be exactly solved by the Bethe-ansatz (BA) method.
The trapped system can be studied in the scheme of the local density approximation
(LDA).

The model (\ref{HF}) is exactly solved by the BA method \cite{C.N.Yang} with the BA
wavefunction
\begin{eqnarray}
 \varphi (x_1,\cdots, x_N)
&=& \sum_Q\sum_P \theta (x_{Q1} \leq \cdots \leq x_{QN}) \times
\nonumber \\
& & [Q,P]\exp [i\sum_{j=1}^Nk_{Pj}x_{Qj}] , \label{2}
\end{eqnarray}
where $k_i$ represent quasimomenta,  $P$ and $Q$ represent permutations of $k_i$ and
$x_i$, respectively.
For the eigenstate with the total spin $S= N/2 - M$ ($M=N_\downarrow$), the coefficient
$[Q,P]$ can be explicitly expressed as $ \lbrack Q,P]=\sum_{T=1}^{C_N^M}\Phi
(y_{T_1},y_{T_2},\cdot \cdot \cdot y_{T_M};P)\prod_{j=1}^M\chi _{y_{T_j},\downarrow
}\prod_{x_i\neq y_{T_j}}\chi _{x_i,\uparrow } $, where $\chi _{x_i,\uparrow }$ ($\chi
_{y_j,\downarrow }$)  denotes
the up (down)-spin, $T$ is a combination of $M$ down-spins in $N$ particles, $\{y_{T_j}\}$ are $M$ elements of $%
T\{x_i\}$, and $ \Phi (y_{T_1},y_{T_2},\cdot \cdot \cdot
y_{T_M};P)=\sum_RA(R)\prod_{j=1}^MF_P(\Lambda _{R_j},y_{T_j}) $
with $R$ being the permutations of $\Lambda s$, $ A(R)= \epsilon
(R)\prod_{j<l}(\Lambda _{R_j}-\Lambda _{R_l}-ic)$, and $
F_P(\Lambda _{R_j},y_{T_j})=
\prod_{j=1}^{y_{T_j}-1}(k_{P_j}-\Lambda _{R_j}+ic/2)
\prod_{l=y_{T_j}+1}^N(k_{P_l}-\Lambda _{R_j}-ic/2)$. The
parameters $k_j$ and $\Lambda _\alpha $ are determined by the
Bethe-ansatz equations (BAEs) \cite{C.N.Yang}:
\begin{eqnarray}
&& \hspace{10mm} k_jL=2\pi I_j-2\sum_{\alpha =1}^M\tan
^{-1}(\frac{k_j-\Lambda
_\alpha }{c/2}), \label{BAE1} \\
&& \sum_{j=1}^N2 \tan ^{-1}(\frac{\Lambda _\alpha -k_j}{c/2})=2\pi
J_\alpha +2\sum_{\beta =1}^M\tan ^{-1}(\frac{\Lambda _\alpha
-\Lambda _\beta }c) .\nonumber\\ \label{BAE2}
\end{eqnarray}
The eigenenergies are given by $E=\frac{\hbar^2}{2m} \sum_j^N k_j^2$. Here both $k_j$
and $\Lambda_\alpha$ are real numbers if $c>0$. The ground state solution corresponds
to $I_j =(N+1)/2-j$ and $J_\alpha =(M+1)/2-\alpha$. In the limit of $cL \gg 1$,
$\Lambda_\alpha$ are proportional to $c$, but $k_j$ remain finite, therefore the
quasi-momenta can be given approximately
\begin{equation}
k_j L = 2\pi I_j-\zeta \frac{k_j}{|c|} + O(|c|^{-3}) \label{KTG}
\end{equation}
with $\zeta=\sum_{\alpha=1}^M \frac 1{(\Lambda _\alpha /c)^2+1/4}$.
It follows that the ground energy in the strongly repulsive limit reads
\begin{equation}
E_{FTG} = \frac{\hbar^2}{2m} \frac{\pi^2}{3L^2}N(N^2-1) (1+\frac{\zeta}{L|c|})^{-2}+ O(|c|^{-3}),
\label{ETG}
\end{equation}
which is consistent with the result in Ref.\cite{GuanXW} up to order of $c^{-1}$. In
the limit of $c\rightarrow\infty$, the ground energy is identical to that of a
polarized N-fermion system.
\begin{figure}[tbp]
\includegraphics[width=9.0cm]{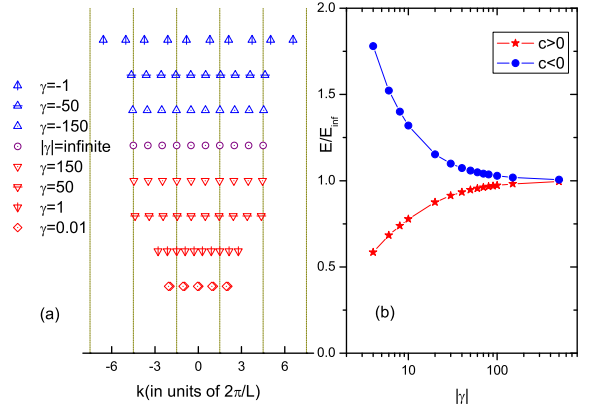}
\caption{ (color online) (a) Quasi-momentum distributions for the ground state of the
repulsive Fermi gas and the FSTG state of the attractive Fermi gas with different
values of $\gamma$. (b) The energies $E_{FTG}$ (stars) and $E_{FSTG}$ (dots) vs
$\gamma$.} \label{momentum}
\end{figure}

\textit{FSTG state.---} If the interaction is attractive, the ground state is composed
of $N-2M $ real $k_i$ and $2M$ complex ones. In the limit $-cL \gg 1$, the complex
solutions take the 2-string form: $ k_\alpha \approx \Lambda _\alpha +\frac c2i$, and
$k_{M+\alpha }\approx \Lambda _\alpha -\frac c2i$.
Except the complex solutions, the BAEs also have real solutions for $c<0$, which,
however, correspond to some highly excited states of attractive Fermi systems. The FSTG
state corresponds to the lowest real solutions of BAEs (\ref{BAE1}) and (\ref{BAE2})
with $c<0$. In this case, $\Lambda_\alpha$ go infinite and $k_j$ remain finite with
$|c|L\rightarrow \infty $, thus the momenta are given by
\begin{equation}
k_jL = 2\pi I_j+\zeta \frac{k_j}{|c|} + O(|c|^{-3}) \text{.}
\label{KSTG}
\end{equation}
Despite the $\zeta$ in Eq. (\ref{KSTG}) having the same form as in Eq. (\ref{KTG}),
generally $\zeta(c) \neq \zeta(-c)$ since the solutions $\Lambda_\alpha$ of Eq.
(\ref{BAE2}) are not symmetric for $c$ and $-c$. However, in the strong coupling limit,
up to order of $c^{-1}$ Eq.(\ref{BAE2}) becomes $ 2N\tan ^{-1}(\frac{\Lambda _\alpha
}{c/2})=2\pi J_\alpha
+2\sum_{\beta =1}^M\tan ^{-1}(\frac{\Lambda _\alpha -\Lambda _\beta }%
c)+O(|c|^{-2})\text{,} $ which is invariant under the operation $P:\{c\rightarrow
-c,\Lambda _\alpha \rightarrow -\Lambda _\alpha \}$. Therefore we have $\zeta(c) =
\zeta(-c)$  up to the order of $c^{-2}$.  The energy of the FSTG gas in the strongly
attractive limit is thus given by
\begin{equation}
E_{FSTG} = \frac{\hbar^2}{2m} \frac{\pi^2}{3L^2}N(N^2-1) (1 - \frac{\zeta}{L|c|})^{-2}+ O(|c|^{-3}).
\label{ESTG}
\end{equation}
In the limit of $|c|\rightarrow \infty$, we have $E_{FSTG} = E_{FTG}$ and $k_j=I_j 2
\pi/L$ for both the Fermi TG and the FSTG gas. In Fig. 1(a), for an example system with
$N=10$ and $M=5$, we show the BAE solutions of $k_j$ for the repulsive Fermi gas and
the attractive FSTG gas with different values of $\gamma=c/\rho$, where $\rho=N/L$ is
the particle density. The quasimomentum distributions for the repulsive Fermi gas and
the STG gas approach the same limit from different sides when $|\gamma|$ goes infinite.
Correspondingly, $E_{FSTG}$ and $E_{FTG}$ also approach the same limit
$E_{\text{inf}}=\frac{\hbar^2}{2m} \frac{\pi^2}{3L^2}N(N^2-1)$ as shown in Fig.1(b).

The FSTG state can be achieved through a similar sudden switch as in Ref.\cite{Haller}.
The system is first prepared at the ground state in the strongly repulsive regime,
i.e., $|\Psi (t=0)\rangle =|\varphi _0(c)\rangle $. After a sudden switch into the
strongly attractive regime with $c^{\prime }<0$, the wavefunction is given by $|\Psi
(t)\rangle =e^{-iHt}|\varphi _0(c)\rangle =\sum_ie^{-iE_it}\alpha _i|\varphi
_i(c^{\prime })\rangle$,
where $\varphi _i(c^{\prime })$ is the i-th eigenstate of the Hamiltonian with
parameter $c^{\prime }$ and $\alpha _i=\langle \varphi _i(c^{\prime })|\varphi
_0(c)\rangle $. The probability for the system staying in a state $|\varphi
_i(c^{\prime })\rangle $ is given by $|\alpha _i|^2=|\langle \varphi _i(c^{\prime
})|\varphi _0(c)\rangle |^2$. We note that the wave-functions $\varphi _{STG}(c^{\prime
})$ and $ \varphi_0(c)$ are identical when $c^{\prime }=-\infty$ and $c = \infty$, and
thus one can expect the probability of the system transforming from the Fermi TG gas to
STG phase to be close to $1$ for large $|c^{\prime }|$ and $|c|$. In Fig 2, we display
transition probabilities from the initial ground state with $c>0$ to the STG phase with
$c^{\prime}=-c$ for different-size systems. Our results show that the overlap of
wave-functions approaches the limit of $1$ for large $|\gamma|$, although it decreases
as the system size increases. In the strongly interacting limit, the transition
probabilities can be approximately represented as $P(N,M,\gamma)=1-a(N,M)/ \gamma^2$.
The upper bound of the parameter $a(N,M)$ can be estimated as $a(N,M) \leq 64\pi^2 N
M^2$ through the expansion of wavefucntions to order of $1/c$. The transition
probability for a larger system is expected to approach $1$ if $|\gamma| \gg M
N^{1/2}$. However, the calculation for a large system becomes a very time-consuming
task due to the calculation of multidimensional integrals.
\begin{figure}[tbp]
\includegraphics[width=7.0cm,angle=-0]{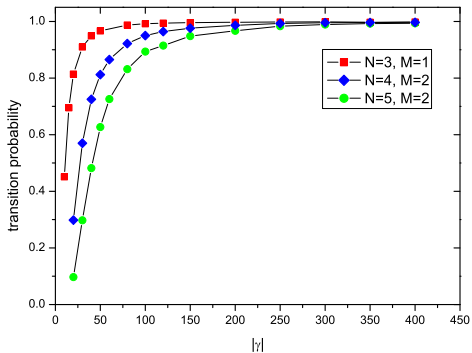}
\caption{ Transition probabilities from the Fermi TG gas to FSTG phase for systems with
$N=3,4,5$. } \label{transition}
\end{figure}
\begin{figure}[tbp]
\includegraphics[width=7.0cm]{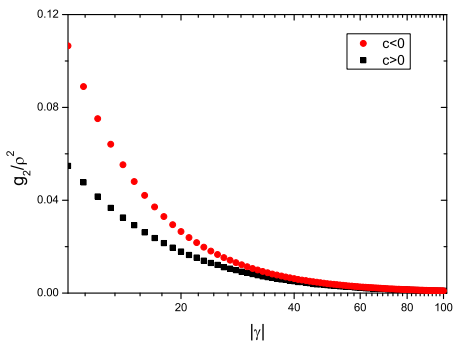}
\caption{(color online) Local correlation energy vs $\gamma$ for the FSTG state and
ground state of the repulsive Fermi gas.} \label{ES}
\end{figure}
\begin{figure}[tbp]
\includegraphics[width=7.0cm]{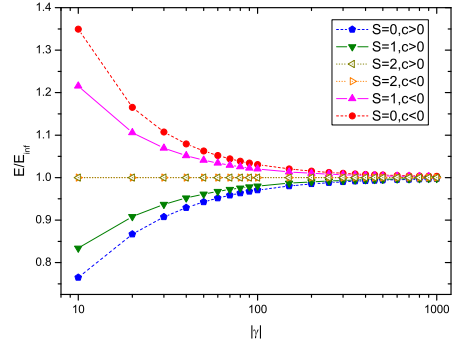}
\caption{Energy vs $\gamma$ for states with different total spin $S$.} \label{ES}
\end{figure}

In thermodynamic limit where $N$, $M$, and $L$ go infinite but $\rho=N/L$ and $m=M/L$
remain finite, the BAEs can be expressed in the form of the coupled integral equations
\cite{C.N.Yang}. The energy per particle reads $ \epsilon (\gamma) =\frac{\hbar
^2}{2m}\rho^2e(\gamma ) $. For the case with $m=\rho/2=\int_{-\infty}^{\infty}
\sigma(\Lambda) d \Lambda$, $\zeta$ can be calculated via the integral $\zeta=
\int_{-\infty}^{\infty} dx 4 \sigma(x) /(x^2+1) =2 \ln 2 $, where
$\sigma(x)=1/[4\cosh(\pi x/2)]$. Therefore we can obtain the energy expansion
$e_{STG}(\gamma)=\frac{{\pi}^2}{3}[ 1+4\ln 2|\gamma| ^{-1}+12(\ln 2)^2|\gamma|
^{-2}]+O(\gamma ^{-3})$ for large $|\gamma|$.
The energy in the whole area can be calculated by numerical solving the coupled
integral equations. Comparing the expansion result with the exact numerical result, we
find that they agree very well in the regime of large $\gamma$, for example,
$|e_{expansion}-e_{num}|/|e_{num}|< 10^{-3}$ for $\gamma=25$. A characteristic of the
STG gas is that it has stronger correlations than the TG gas. In the regime of $
|\gamma |\gg 1$, the local two-particle correlation function $g_2(\gamma
)=\rho^2de(\gamma )/d\gamma $ can be directly derived from $e(\gamma)$, which gives $
g_2(\gamma)_{TG}/\rho^2 \approx (4\pi^2/3)(\ln 2 |\gamma |^{-2}-6(\ln 2)^2|\gamma
|^{-3}) $, and $ g_2(\gamma)_{STG}/\rho^2 \approx (4\pi^2/3) (\ln 2 |\gamma |^{-2}+
6(\ln 2)^2|\gamma |^{-3}) $. Thus we have $g_2(-|\gamma |)_{STG}>g_2(|\gamma |)_{TG}$
as shown in Fig.3.

In terms of terminologies of Tomonaga-Luttinger liquid (TLL) theory \cite{Recati}, the
strongly repulsive phase of the spin-balanced Fermi gas corresponds to a TLL with the
charge TLL parameter $K_c \approx (1+4 \ln 2/|\gamma|)/2>1/2$ \cite{Recati}. The FSTG
phase corresponds to a highly excited gas-like state where unpaired particles are
strongly correlated. This strongly collective behavior may be phenomenologically
described by  $K_c \approx (1-4 \ln 2/|\gamma|)/2$ in the strongly interacting limit,
which is smaller than $1/2$. For both cases, the spin TLL parameter $K_{\sigma}=1$ due
to spin-rotational invariance.

\textit{Degeneracy of the FSTG state.---} In comparison to the Bose system, the ground
state of a spin-1/2 system is highly degenerate in the TG limit due to the fact that
states with different total spins have the same energy \cite{Guan}. However, for a
large but finite interaction strength the degeneracy is broken and the true ground
state is the state with the lowest $S$. For the spin-1/2 system described by
(\ref{HF}), one can understand this fact from the energy expression (\ref{ETG}). The
term $\zeta$ is $M$-dependent $(M=N/2-S)$ and we have $\zeta(M_1) < \zeta(M_2)$ if $M_1
< M_2$, which leads to $ E_{FTG}(S_2) < E_{FTG}(S_1)$ for $S_2 < S_1$. This is
consistent with the Lieb-Mattis theorem \cite{Lieb-Mattis}. The energy difference is
proportional to $1/c$
and vanishes as $c\rightarrow \infty$. On the other hand, for the FSTG state we have $
E_{FSTG}(S_2) > E_{FSTG}(S_1)$ for $S_2 < S_1$ according to Eq. (\ref{ESTG}), {\it
i.e.}, the FSTG state with the smaller S has higher energy. To give an example, we
calculate the ground state energy and the energy of the FSTG state for a system with
$N_{\uparrow} =N_{\downarrow}=2$. As shown in Fig {\ref{ES}}, energies for states with
different total spins approach the same limit of the polarized Fermi gas as
$|c|\rightarrow \infty$.

\textit{Experimental detection.---} To realize the FSTG gas, one can first tune the
interaction of the 1D Fermi gas to the strongly repulsive regime by the Feshbach
resonance \cite{Moritz}, and then suddenly switch the interaction across the resonance
point. Similar to the bosonic case, one can measure the frequency of the breath mode of
the FSTG gas subjected to a weak harmonic confinement along the axial direction, which
is sensitive to various regimes of interaction. For the Fermi gas in a harmonic trap
with $V_{ext} = m \omega_x^2 x^2/2$, we can determine the density distribution of the
STG gas within the LDA. According to the LDA, the system is in local equilibrium at
each point $x$ in the external trap. The density distribution of the FSTG gas is then
obtained via the local equation of state $\mu_0=\mu[\rho(x)] + V_{ext}(x)$
\cite{Dunjko,Menotti,Astrakharchik3}. Here $\mu(\rho)=
\partial_{\rho}[\rho \epsilon(\rho) ]$ is the local chemical potential with
$\epsilon(\rho)$ the energy density of the homogenous FSTG gas to be determined by
numerically solving integral BA equations,
and $\mu_0$ is determined by the normalization condition $\int dx \rho(x)=N$. Following
Refs. \cite{Menotti,Astrakharchik3}, we calculate the frequency of the lowest breathing
mode from the mean square radius of the trapped FSTG gas via $ \omega^2=-2 \langle x^2
\rangle/(d\langle x^2 \rangle/d\omega_x^2)$ with $\langle x^2 \rangle = \int \rho(x)
x^2 dx/N$. The solid line in Fig.5 shows the frequency of breathing mode of the
attractive FSTG gas as a function of the interaction strength $N a_{1d}^2/a_x^2$ with
the harmonic oscillator length $a_x=\sqrt{\hbar/m\omega_x}$. The frequency of the
breath mode for the FSTG gas exhibits a peak with a maximum of $\omega^2/\omega_x^2$
about $4.3$. We also give results of the repulsive Fermi TG gas (the dashed line) and
the ground state of the attractive Fermi gas (the dotted line) for comparison. These
results are essentially based on the sum-rule approach and provide generally an upper
bound on the frequencies \cite{Menotti}.
\begin{figure}[tbp]
\includegraphics[width=8.00cm]{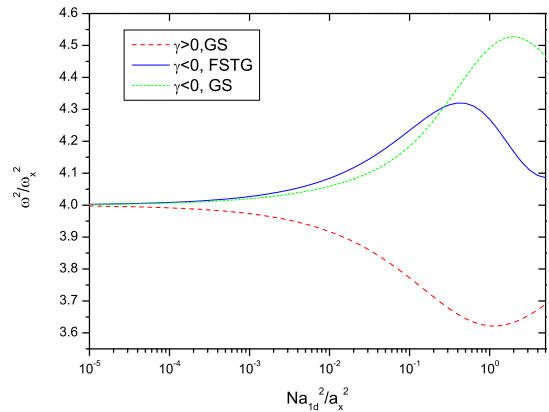}
\caption{Square of the lowest breathing mode frequency vs the interaction strength $N
a_{1d}^2/a_x^2$ for the FSTG gas (solid line), the repulsive Fermi TG gas (dashed line)
and the ground state of the attractive Fermi gas (dotted line). \label{breathing mode}}
\end{figure}

\textit{Summary.---} In summary, we study the realization of the FSTG phase in the
interacting spin-1/2 Fermi gas. Starting from the ground state of a strongly repulsive
Fermi gas, the FSTG state can be realized by a sudden switch of interaction to the
strongly attractive regime. It is shown that the FSTG state is stable against forming
pairing states even in the presence of the strongly attractive interaction between
fermionic atoms with opposite spins. We also calculate the lowest breathing mode
frequency of the FSTG gas which may be detected by the experiment.

\textit{Note added.---} While this work was being prepared for submission, a related
manuscript appeared \cite{Girardeau_FSTG}, in which the FSTG gas in a harmonic trap is
studied.

We thank X.-W. Guan and X. Yin for helpful discussion. This work has been supported by
the NSF of China under Grants No. 10974234 and No. 10821403, 973 grant and National
Program for Basic Research of MOST.


\begin{thebibliography}{9}
\bibitem{Bloch} I. Bloch, {\it et.
al.}, Rev. Mod. Phys. \textbf{80}, 885 (2008).

\bibitem{Lewenstein} M. Lewenstein {\it et.
al.}, Adv. Phys. \textbf{56}, 243 (2007).

\bibitem{Greiner} M. Greiner \emph{et al.},  Nature \textbf{415}, 39
(2002).

\bibitem{Paredes}  B. Paredes, {\it et. al.}, Nature \textbf{429}, 277 (2004).


\bibitem{Kinoshita}  T. Kinoshita, {\it et. al.}, Science \textbf{
\ 305}, 1125 (2004).

\bibitem{Haller}  E. Haller, {\it et. al.}, Science \textbf{325}, 1224
(2009).


\bibitem{Astrakharchik1} G. E. Astrakharchik, J. Boronat, J. Casulleras, and
S. Giorgini, Phys. Rev. Lett. \textbf{95}, 190407 (2005).

\bibitem{Astrakharchik2} G. E. Astrakharchik, D. Blume , S. Giorgini and B.
E. Granger, Phys. Rev. Lett. \textbf{92}, 030402 (2004).

\bibitem{Batchelor} M. T. Batchelor, M. Bortz, X. W. Guan, N. Oelkers, J.
Stat. Mech. (2005) L10001.

\bibitem{Tempfli} E. Tempfli, {\it et al.}, New J. Phys. 10, 103021 (2008).


\bibitem{Chen} S. Chen, {\it et al.}, Phys. Rev. A \textbf{81}, 031609(R) (2010).

\bibitem{Girardeau_STG} M. D. Girardeau and G. E. Astrakharchik, Phys. Rev. A \textbf{81},
061601(R) (2010).

\bibitem{Moritz} H. Moritz, {\it et al.}, Phys. Rev. Lett. \textbf{94}, 210401 (2005).

\bibitem{C.N.Yang} C. N. Yang, Phys. Rev. Lett. \textbf{19}, 1312 (1967).

\bibitem{Gaudin}  M. Gaudin, Phys. Lett. \textbf{24A}, 55 (1967).



\bibitem{Chen2} S. Chen, {\it et al.}, Phys. Rev. A \textbf{81}, 031608(R) (2010).

\bibitem{Olshanii}  M. Olshanii, Phys. Rev. Lett. \textbf{81}, 938 (1998).

\bibitem{Girardeau} M. D. Girardeau, J. Math. Phys. \textbf{1}, 516 (1960).

\bibitem{Girardeau07} M. D. Girardeau and A. Minguizzi, Phys. Rev. Lett. 99, 230402
(2007).


\bibitem{Guan} L. Guan, {\it et al.}, Phys. Rev. Lett. \textbf{102},
160402 (2009).

\bibitem{GuanXW} N. Oelkers, M. T. Batchelor, M. Bortz and X.-W. Guan,
J. Phys. A: Math. Gen. \textbf{39},  1073 (2006).

\bibitem{Recati} A. Recati, P. O. Fedichev, W. Zwerger, and P. Zoller,
Phys. Rev. Lett. \textbf{90}, 020401 (2003); J. Opt. B \textbf{5}, S55 (2003).

\bibitem{Lieb-Mattis} E. H. Lieb and D. Mattis, Phys. Rev. \textbf{125}, 164 (1962).

\bibitem{Dunjko} V. Dunjko, V. Lorent, and M. Olshanii, Phys. Rev. Lett. {\bf 86}, 5413 (2001).



\bibitem{Menotti} C. Menotti and S. Stringari, Phys. Rev. A \textbf{66},
043610 (2002).


\bibitem{Astrakharchik3} G. E. Astrakharchik, D. Blume, S. Giorgini, and L. P. Pitaevskii,
Phys. Rev. Lett. \textbf{93}, 050402 (2004).



\bibitem{Girardeau_FSTG} M. D. Girardeau, arXiv:1004.2925.


\end{thebibliography}
\end{document}